\documentclass[aps,twocolumn,superscriptaddress,showkeys]{revtex4}   %%preprint

    \newcommand{\re}{\mathop{\mathrm{Re}}}
    \newcommand{\im}{\mathop{\mathrm{Im}}}

    \newcommand{\ThreeJSymbol}[6] {\ensuremath{
                 \left( \begin{array}{ccc}
                  {#1}&{#2}&{#3} \\
                  {#4}&{#5}&{#6}
                 \end{array} \right)
                                  }}

\usepackage{amsmath}
\usepackage{graphicx}
\usepackage{epsfig}

\usepackage{color}
\usepackage{bm}

\begin{document}

\title{Xclaim: a graphical interface for the calculation of core-hole spectroscopies}

%%%%%%%%%%%%%%%%%% 

\author{Javier Fern\'{a}ndez-Rodr\'{\i}guez}

\affiliation{Department of Physics, Northern Illinois University, DeKalb, Illinois 60115, USA}
\affiliation{Advanced Photon Source, Argonne National Laboratory, 9700 South Cass Avenue, Argonne, Illinois 60439, USA}

\author{Brian Toby}

\affiliation{Advanced Photon Source, Argonne National Laboratory, 9700 South Cass Avenue, Argonne, Illinois 60439, USA}

\author{Michel van Veenendaal}

\affiliation{Department of Physics, Northern Illinois University, DeKalb, Illinois 60115, USA}
\affiliation{Advanced Photon Source, Argonne National Laboratory, 9700 South Cass Avenue, Argonne, Illinois 60439, USA}

\date{\today}

\begin{abstract}

Xclaim (x-ray core level atomic multiplets) is a graphical interface
for the calculation of core-hole
spectroscopy and ground state properties within a charge-transfer
multiplet model taking into account a many-body Hamiltonian with Coulomb,
spin-orbit, crystal-field, and hybridization interactions.
Using   Coulomb and spin-orbit
parameters calculated in the Hartree-Fock limit and ligand field parameters (crystal-field, hybridization
and charge-transfer energy)
the program calculates x-ray absorption spectroscopy (XAS), x-ray photoemission spectroscopy (XPS),
photoemission spectroscopy (PES) and inverse photoemission (IPES).
The program runs on Linux, Windows and MacOS platforms.

\end{abstract}

%\pacs{78.70Dm,71.27.+a,71.70.Ch}
%http://ufn.ru/en/pacs/
%78.70Dm XAS
%71.27.+a Strongly correlated electron systems; heavy fermions
%71.70.Ch Crystal and ligand fields


\maketitle

\section{Introduction}

Multiplet ligand-field theory (MLFT) or small cluster calculations~\cite{Ballhausen1962,deGrootKotani,vdL_guide_2006,vanVeenendaal2015book}
are useful approaches for calculating x-ray spectroscopy on strongly correlated materials where the spectral lineshape is dominated by strong multiplet effects arising from the Coulomb interactions between the valence electrons and between the valence electrons and the core hole. Since the eigenfunction of a Coulomb multiplet often involves several Slater determinants, they are often poorly described by effective single-particle models, such as density functional theory. For MLFT, one considers a single ion and the effects of the ligands are described by an effective crystal field. This approach often works well when describing x-ray absorption spectroscopy. For x-ray photoemission, screening effects are stronger and the ligands have to be included explicitly. This is generally known as small-cluster calculations.
The spectra are calculated by constructing a many-body Hamiltonian for the system
 using full configuration-interaction, {\it i.e.}  
taking into account in the basis states any possible combination of
Slater determinants and has the advantage of accurately  treating
the Coulomb interaction in the metal ion.
This approach has been used with great success to describe
x-ray spectra.~\cite{Thole1985PRB,deGroot1990,TanakaJo1994,vanderLaan1991,missingURL,Stavitski2010,Mirone2008,HilbertManual,Uldry2012,multixURL,jfrFePc2014}.

In this paper we discuss the calculation of core-hole spectroscopy in
terms of a multiplet Hamiltonian with the model implemented
in Xclaim~\cite{xclaimURL}. Xclaim is a code to calculate different types of X-ray spectra within the ionic or small-cluster limit. The program allows flexible input and output via a graphical user interface. 
The paper is outlined as follows. First, we give an overview of the model Hamiltonian used in the calculation of the spectra. We discuss the different interactions included in
the many-body Hamiltonian. Subsequently, we describe the various spectroscopies that can be calculated:
x-ray absorption spectroscopy (XAS), x-ray photoemission spectroscopy (XPS),
photoemission spectroscopy (PES) and inverse photoemission (IPES).
The final Section contains a description of  the graphical interface for the calculation of
spectra and ground state properties.

\section{Model Hamiltonian}
The Hamiltonian for MLFT and small-cluster calculations, can be split into the following terms
\begin{equation}
H = H_{\mathrm{atomic}}+H_{\mathrm{CF}}+H_{\mathrm{hybridization}},
\label{Hamtot}
\end{equation}
where $H_{\mathrm{atomic}}$ describes the central ion where the x-ray transition takes place. The last two terms on the right-hand side describe the effects of the surrounding ions: $H_{\mathrm{CF}}$  describes the effects of the ligands as an effective point-charge crystal field; $H_{\mathrm{hybridization}}$ is included for small-cluster calculations and describes the hybridization of the central ion with the nearest-neightbor ligand ions.

The  Hamiltonian for the electrons on the central ion is given by
\begin{eqnarray}
H_{\mathrm{atomic}} &=& \sum_i \frac{\mathbf{p}_i^2}{2m} - \sum_i \frac{Ze^2}{r_i}
+\sum_{i<j}\frac{e^2}{|\mathbf{r}_{i}-\mathbf{r}_{j}|} 
 \nonumber \\
  && +\sum_{i} \zeta(r_i)\mathbf{l}_i \cdot \mathbf{s}_i -\bm{\mu}\cdot \mathbf{B},
\label{Hamiltonianat}
\end{eqnarray}
where the indices $i$ and $j$ run over all  electrons of the ion.
The first term in $H_{\mathrm{atomic}}$ is the kinetic energy, where $\mathbf{p}_i$ and $m$ are the momentum and the mass of the electrons, respectively. The second term on the right-hand side is  the potential energy of the nucleus, where $Z$ is the atomic number. These two terms lead to the binding energy for the electrons. The next term is the Coulomb interaction between the electrons. The  Coulomb interactions that include electrons in a closed shell lead to an effective change in the binding energy. The interaction between two electrons that are both in open shells leads to  multiplet structure that can often be clearly observed in the spectral line shape. The fourth term is  the spin-orbit interaction, where $\zeta(r_i)$ is the radial part of the spin-orbit interaction. The last term is an external magnetic field $\mathbf{B}$ with $\bm{\mu}$ is the total magnetic moment
$\bm{\mu}=-\mu_{B}(\mathbf{L}+g_S\mathbf{S})$ with $g_S\approx 2$ the spin gyromagnetic ratio.The interaction is weak but plays a crucial role in lifting the degeneracy of the ground state for magnetic systems.  Let us consider the interactions in $H$ in more detail. 

\subsection{Coulomb and spin-orbit interactions}

In the evaluation of the matrix elements of the Hamiltonian, the angular part, resulting from the integrals over the spherical harmonics of the atomic wavefunctions, can be expressed in terms of Clebsch-Gordan coefficients or $3j$ symbols. The radial parts of the matrix elements involve integrals over the radial  atomic wavefunctions and therefore explicitly depend on the effective central-field potential that an electron in a particular orbital experiences.  Whereas the angular matrix elements can be evaluated analytically, the radial matrix elements need to be calculated numerically. In the calculations, the Hartree-Fock self-consistent atomic field for an isolated ion, as implemented in Cowan's atomic multiplet program
RCN~\cite{CowanBook,CowanURL}, is used. The resulting radial wavefunctions $P_{nl}(r)$, where $n$ is the principal quantum number and $l$ is the angular momentum quantum number, can be used to evaluate the matrix elements.

For the spin-orbit interaction this gives 
\begin{equation}
\zeta_{nl}=\frac{\alpha^2}{2}\int^{\infty}_0 \frac{1}{r} \left( \frac{dU_{nl}}{dr}\right) |P_{nl}(r)|^2 r^2dr,
\end{equation}
 $\alpha$ is the fine structure constant. 
In the matrix element the radial part of the interaction $\zeta(r_i)$ from Eq. (\ref{Hamiltonianat}) can be expressed in terms of the derivative of the effective central-field potential energy $U_{nl}$.
The spin-orbit interaction  mixes the orbital and spin quantum numbers. This can change the ground-state symmetry, which can significantly alter the spectral line shape. In the final state, 
the spin-orbit of a core-shell with $l_c>0$ is often large enough to separate the spectrum into two distinct edges,  for example the $L_2$~($2p_{1/2}$) and $L_3$~($2p_{3/2}$) edges for transition-metal compounds.

For the electron-electron interaction, we can make a multipole expansion of $1/r_{ij}$ with $r_{ij}=|{\bf r}_i-{\bf r}_j|$
\begin{equation}
\label{DefHCoulomb}
\frac{e^2}{|\mathbf{r}_{i}-\mathbf{r}_{j}|} = e^2 \sum_{k} \frac{r^k_{<}}{r^{k+1}_{>}} \mathbf{C}^k({\hat {\bf r}}_i) \cdot \mathbf{C}^k({\hat {\bf r}}_j)  ,
\end{equation}
where $r_{<,>}$ is the lesser/greater of $r_i$ and $r_j$; $\mathbf{C}^{k}$ is a tensor of renormalized spherical harmonics whose components are related to the spherical harmonics $C^{k}_q = \sqrt{\frac{4\pi}{2k+1}}Y_{kq}$;
${\hat {\bf r}}={\bf r}/r$ is a shorthand for the angular coordinates $\theta$ and $\varphi$ in spherical polar coordinates.  The Coulomb interaction is customarily parametrized in terms of the radial integrals
\begin{eqnarray}
&&R^k_{n_1l_1n_2l_2n_3l_3n_4l_4}= e^2 \int^\infty_0 dr_1 \, r_1^2  \int^\infty_0 dr_2 \, r_2^2 \frac{2r^k_{<}}{r^{k+1}_{>}}     \nonumber \\
&&~~~~~~~~~~~ P_{n_4l_4}(r_1)P_{n_3l_3}(r_2)P_{n_2l_2}(r_2)  P_{n_1l_1}(r_1),
\end{eqnarray}
For two atomic orbitals, the integrals 
are divided into direct $F^k_{nl;n'l'}=R^k_{nln'l'n'l'nl}$ and exchange $G^k_{nl;n'l'}=R^k_{nln'l'nln'l'}$ parts. In addition, one has $F^k_{nl;nl}=G^k_{nl;nl}$.
The matrix element for the Coulomb interaction between two electrons with orbital angular momentum $l$ and $l'$ can be written in 
\begin{eqnarray}
&& \langle nl m_4 \sigma; n'l' m_3\sigma'|H| n'l'm_2 \sigma';n lm_1\sigma\rangle = \nonumber   \\
&& \quad \left [
 \sum_{kq} F^k_{nl;n'l'}\langle lm_4|C^{k}_q | lm_1\rangle  \langle l'm_3|[C^{k}_q]^* | l'm_2  \rangle 
\right . \\
&& \quad \left . - \delta_{\sigma\sigma'} G^k_{nl;n'l'}  \langle l'm_3|C^{k}_q | lm_1 \rangle   \langle lm_4|[C^{k}_q ]^*| lm_2 \rangle  \right ] \nonumber 
\end{eqnarray}
where the matrix elements of the normalized spherical harmonics $C^{k}_q$ are given by
\begin{equation}
\label{MECkq}
\langle l'm'|C^{k}_q | lm \rangle  = (-1)^m [ll']^{1/2} \ThreeJSymbol{l'}{k}{l}{0}{0}{0} \ThreeJSymbol{l'}{k}{l}{-m'}{q}{m}
\end{equation}
and $[ll']$ denotes $(2l+1)(2l'+1)$.

For ionic materials, the Hartree-Fock values of $F^k$ and $G^k$ are reduced customarily to 80\%
to account for the intraatomic configuration-interaction effects.
Reductions to less than 80\% can be used to mimic the effect of
hybridization in a simpler crystal field model without ligand orbitals.~\cite{ZaanenWestraSawatzky1986}
The reduction in the Coulomb parameters is related to an increase
in hybridization and indicates a decrease in
importance of the core-valence interaction.  A strong reduction
can occur when the excitonic final states in XAS are not
completely pulled below the valence band continuum.
Reductions of about 50\% are necessary for shallow core-hole edges as
the $M_{2,3}$ (3s) edges of transition
metals~\cite{Berlasso2006PRB,Wray2012} and the $O_{4,5}$ (5d) edges of
actinides~\cite{Bradley2010PRB}.

\subsection{Crystal Field}

The spectral lineshape is generally strongly affected by solid-state
effects. To lowest order, these effects can be included by an
effective crystal field, $H_{\mathrm{CF}}$ in Eq. (\ref{Hamtot}). This
not only describes the point-charge crystal field, but can often also
account for some of the effects of the hybridization of the central
ion with the surrounding ligands. The effect of the crystal field is
to lower the symmetry causing a splitting of the states that are
obtained in spherical symmetry, {\it i.e.} by including only Coulomb
and spin-orbit interactions. The spectral lineshape changes because of
the energy splittings caused by the crystal field and due to the
change in the symmetry of the ground state.

Many conventions exist for parametrizing the effect of the ligand
environment. In our code, for the crystal field we use a
parametrization based on the point group of the ion in terms of
Ballhausen or Wybourne parameters~\cite{Mulak2000,Haverkort2005}.  For
a shell with angular quantum number $l$, the crystal field is written
in terms of spherical harmonics as
\begin{equation}
\label{CFham}
   H_{\mathrm{CF}}= \sum_{k,q} B_{kq}C^{k}_q
\end{equation}
with $0\leq k\leq 2l$, $k$~an even integer and $-k\leq q \leq k$; 
$B_{kq}$  are the Wybourne crystal-field parameters.

The hermiticity of the Hamiltonian imposes $B_{k,-q}= (-1)^q B^{*}_{kq}$. By separating the real and imaginary parts of the Wybourne parameters $B_{kq}=\re B_{kq}+i\im B_{kq}$, we can rewrite
$ H_{\mathrm{CF}}$  as
\begin{eqnarray}
  H_{\mathrm{CF}}&=&  \sum_k \Big\{ B_{k0} C^{k}_0 +\nonumber\\
&& \sum_{1\leq q \leq k} \Big[ \re B_{kq} \Big(  C^{k}_q +(-1)^q C^{k}_{-q} \Big)  \nonumber\\
&&~~~~~+ i  \im B_{kq}  \Big(  C^{k}_q -(-1)^q C^{k}_{-q}   \Big) \Big] \Big\}
\end{eqnarray}
The number of non-zero parameters, and algebraic relationships between them are determined by the
point group symmetry.
The Wybourne parameters $B_{kq}$ can be easily related to the Stevens
parameters~\cite{CFHandbook,Rotter2012JPCM,McPhaseManual}.

In the case of cubic octahedral symmetry $(O_h)$, the 4-fold axis around $z$ limits the 
values of $q$ to $q=0,\pm 4$,
and the invariance of the crystal field under $90$-degree rotations about $x$ or $y$ forbids 
the term $B_{20}$, since the spherical harmonic $C^{2}_0$ is not preserved for those rotations,
and relates the $q=0$ and $q=4$ parameters for the $k=4,6$ representations:
\begin{eqnarray}
  B_{44}&=&\sqrt{\frac{5}{14}}  B_{40}         \nonumber\\
  B_{64}&=&-\sqrt{\frac{7}{2}}  B_{60}  
\end{eqnarray}
The only free parameters are $B_{40}$ and $B_{60}$.  The crystal field Hamiltonian becomes
\begin{eqnarray}
  H_{\mathrm{CF},O_h}&=& 
B_{40} \Big[C^{4}_0  + \sqrt{\frac{5}{14}} \big(C^{4}_{4} + C^{4}_{-4}\big)  \Big] \nonumber \\
&& + B_{60} \Big[ C^{6}_0  -\sqrt{\frac{7}{2}} \big(C^{6}_{4} + C^{6}_{-4}\big) \Big]
\end{eqnarray}
Another common notation in the cubic case is to use the parameters $V_4=B_{40}/8$ and $V_6=B_{60}/16$.~\cite{Magnani2007}
For an $f$-shell, the orbitals split into three independent representations:
$a_{2g}$ ($f_{xyz}$), $t_{1g}$ ($f_{x^3}$, $f_{y^3}$, $f_{z^3}$), and $t_{2g}$
($f_{x(y^2-z^2)}$, $f_{y(x^2-z^2)}$, $f_{z(x^2-y^2)}$) with energies 
$$\varepsilon_{a_{2g}}=\frac{80}{143}B_{60}-\frac{4}{11}B_{40}$$
$$\varepsilon_{t_{1g}}=\frac{100}{429}B_{60}+\frac{2}{11}B_{40}$$
$$\varepsilon_{t_{2g}}=-\frac{60}{143}B_{60}-\frac{2}{33}B_{40}$$
%respectively.
%
For a $d$-shell ($l=2$) the term $B_{60}$ does not contribute, and 
the orbitals split between $e_g$ and $t_{2g}$ orbitals
at energies $\frac{2}{7}B_{40}$ and $-\frac{4}{21}B_{40}$
The parameter $B_{40}$ can be related to the commonly used $10Dq$ parameter for
octahedral splitting by $10Dq=\frac{10}{21}B_{40}$.
We can generalize the splitting in a $d$-shell to tetragonal symmetry
and relate $B_{20}$, $B_{40}$ and $B_{44}$, 
to the parameters $Dq$, $Ds$ and $Dt$~\cite{Ballhausen1962,YangWei2005},
\begin{eqnarray}
  B_{20}&=&-7 Ds                             \nonumber\\
  B_{40}&=&21( Dq -Dt)                       \nonumber\\
  B_{44}&=&21\sqrt{\frac{5}{14}}Dq
\end{eqnarray}

The splitting of the valence shell orbitals is determined by the
point-symmetry group of the crystalline environment.  By making a
unitary transformation to a symmetry-adapted basis it is always
possible to write the crystal-field Hamiltonian as a sum over
irreducible representations,
$$ H_{\mathrm{CF}}= \sum_{\Gamma\gamma } \varepsilon(\Gamma)  c^{\dagger}_{\gamma}c_{\gamma}$$
where $\varepsilon_{\Gamma}$ is the energy of the $\Gamma$ representation of the point group, 
and $c^{\dagger}_{\gamma}$ is the creation operator for an electron in
the $\gamma$ orbital belonging to the $\Gamma$ representation. In the case of cctahedral Symmetry ($O_h$) for a $d$-shell, the $e_g$ ($d_{3z^2-r^2}$, $d_{x^2-y^2}$) and $t_{2g}$ ($d_{xy}$, $d_{yz}$, $d_{xz}$)
orbitals are separated by an energy $10Dq$.
For tetragonal symmetry~($D_{4h}$) the crystal-field splittings of a
$d$-shell orbitals are usually given in terms of the parameters $Dq$,
$Ds$ and $Dt$ (see table~\ref{TablaCFD4h}).

% $\gamma$ denotes the elements of the basis of the $\Gamma$ representation

\begin{table}
\caption{\label{TablaCFD4h} Crystal-field splitting energies for the different irreducible representation $\Gamma$ with components $\gamma$ for $l=2$ in tetragonal symmetry ($D_{4h}$). }
\begin{center}
\begin{tabular}{lcc}
\hline
\hspace{0.3cm}$\Gamma$ \hspace{0.8cm}&   $\gamma$    & $\varepsilon_{\Gamma}$ \\
\hline
\hspace{0.3cm}$a_{1g}$\hspace{0.5cm}  & $3z^2-r^2$ &\hspace{0.8cm} $ 6Dq-2Ds-6Dt$ \hspace{0.8cm} \\
\hspace{0.3cm}$b_{1g}$\hspace{0.5cm} & $x^2-y^2$ & $ 6Dq+2Ds- Dt$   \\
\hspace{0.3cm}$b_{2g}$\hspace{0.5cm}   & $xy$& $-4Dq+2Ds- Dt$  \\
\hspace{0.3cm}$e_{g }$\hspace{0.5cm} & $yz$, $zx$  & $-4Dq- Ds+4Dt$   \\
\hline
\end{tabular}
\end{center}
\end{table}

\subsection{Hybridization}
In many cases, the inclusion of an effective crystal field can lead to a satisfactory interpretation of the spectral line shape, in particular for x-ray absorption spectroscopy. However, when one is dealing with strongly-covalent materials or when interpreting x-ray photoemission, such an approach is inadequate and the ligands need to be included explicitly. The final term $H_{\mathrm{hybridization}}$ in the Hamiltonian in Eq. (\ref{Hamtot}) describes the hybridization of the central ion with the surrounding ligands. Although the number of ligand orbitals is large, the ion only hydridizes with particular symmetry combinations of orbitals. For example, a $d$ orbital only hybridizes with five  linear combinations of ligand orbitals that have the same symmetry properties as the $d$ orbitals. This can be included in the model by an additional shell of effective ligands and
take into account
configurations $d^{n}$, $d^{n+1} \underline{L}^1$,
$d^{n+2} \underline{L}^2\ldots$ where $\underline{L}^n$
denotes $n$ holes in the ligand shell.
Including additional configurations in the model  increases
the computational cost of constructing and diagonalizing our Hamiltonian because of the
increase in the size of the Hilbert space.
We include in the Hamiltonian a hybridization term that mixes the
valence orbitals with an effective ligand shell $L$ with the same number
of orbitals as the valence shell.
We consider only the linear combinations of orbitals for a particular
point symmetry group that couple to the valence shell (i.e.  they
belong to the same irreducible representation as the valence shell
orbitals)

The hybridization term is written as
\begin{equation}
\label{HybHam}
H_{\mathrm{hyb}} =  \sum_{\Gamma\gamma}  T_{dL}(\Gamma) (d^{\dagger}_{\gamma} L_{\gamma} + 
L^{\dagger}_{\gamma} d_{\gamma} ) +\sum_{\Gamma\gamma} \varepsilon_L(\Gamma) L_{\gamma}^{\dagger}L_{\gamma}
\end{equation}
$\varepsilon_L(\Gamma)$ is the on-site energy for the electrons in the ligand shell
depending on the irreducible representation $\Gamma$ to which the ligand orbital $\gamma$ belongs to.
This displacement is produced by the hybridization between the valence orbitals.
$d^{\dagger}_{\gamma}$ and $L^{\dagger}_{\gamma}$ are the creation operators of an
electron in the $d$ and ligand shells.
The transfer integrals $T_{dL}(\Gamma)$ are written in terms of the Slater-Koster
parameters~\cite{SlaterKoster}: $(pd\sigma)$, $(pd\pi)$, related to
the overlap between the $d$ and $p$ orbitals of the ligands. 
An additional term $T_{pp}=(pp\sigma)-(pp\pi)$ splits the $e_g$
and $t_{2g}$ ligand effective orbitals ($T_{pp}$ is approximately $\frac{1}{4}$
of the width of the ligand band).

The transfer integrals and ligand field splittings for octahedral 6-coordinated (MO$_6$) and planar 4-coordinated (MO$_4$) clusters,
in the case of a $d$~metal-center and $p$~ligand orbitals~\cite{EskesPRB} are shown
in table~\ref{TablaHybridizTMO64}.
When setting the values of the Slater-Koster parameters, 
$(pd\pi)$ is expected~\cite{Mattheiss1972} to be slightly less
than $(pd\sigma)/2$.
When considering changes in bond-length, we can use Harrison's
relationship~\cite{HarrisonBook}, i.e. that the $3d$--$2p$ and $2p$--$2p$ charge
transfer integrals are proportional to the power $-3.5$
and $-2$ of the bond-distance, respectively.

\begin{table}
\caption{\label{TablaHybridizTMO64} Hybridization parameters for the TMO$_6$ and TMO$_4$ clusters in the case of a $d$~metal-center and $p$~ligand orbitals
for the group representations in $O_h$ and $D_{4h}$ symmetry in terms of the Slater-Koster
integrals $pd\sigma$, $pd\pi$.}
\begin{center}
\begin{tabular}{lll}
\hspace{0.3cm}$\Gamma$ & $T(\Gamma)$ \hspace{0.6cm}& $\varepsilon_L(\Gamma)$ \hspace{0.3cm}\\
\hline
\hspace{0.3cm}TMO$_6$ $(O_h)$ \hspace{0.6cm} && \hspace{0.3cm}\\
\hline
\hspace{0.3cm}$e_g$    &$\sqrt{3} (pd\sigma)$ \hspace{0.6cm}& $\Delta +T_{pp}$\hspace{0.3cm}\\
\hspace{0.3cm}$t_{2g}$ &$ -2 (pd\pi)     $ \hspace{0.6cm}& $\Delta -T_{pp}$\hspace{0.3cm}\\
\hline
\hspace{0.3cm}TMO$_4$ ($D_{4h}$)\hspace{0.6cm} && \hspace{0.3cm}\\
\hline
\hspace{0.3cm}$a_{1g}$ & $(pd\sigma)          $\hspace{0.6cm} & $\Delta +T_{pp}$ \hspace{0.3cm}\\
\hspace{0.3cm}$b_{1g}$ & $\sqrt{3} (pd\sigma) $\hspace{0.6cm} & $\Delta -T_{pp}$ \hspace{0.3cm}\\
\hspace{0.3cm}$b_{2g}$ & $-2(pd\pi)        $\hspace{0.6cm} & $\Delta +T_{pp}$ \hspace{0.3cm}\\
\hspace{0.3cm}$e_g$    & $-\sqrt{2} (pd\pi)$\hspace{0.6cm} & $\Delta     $ 
\end{tabular}
\end{center}
\end{table}

The two parameters that determine the amount of covalent mixing between the valence shell of the metal-center
and the ligands are the transfer integrals $T$ in Eq. (\ref{HybHam}) and the charge transfer energy $\Delta$.
We define $\Delta$ as the lowest cost in energy of removing one electron
from the ligands and transferring it to the metal center, i.e., the difference between the
lowest eigenenergies for the $d^n$ and $d^{n+1}\underline{L}$ configurations,
\begin{equation}
\Delta = E(d^{n+1} \underline{L})-E(d^{n})  .
\end{equation}
In the absence of hybridization, and neglecting the multiplet splitting,
the total energy for a particular number of electrons can be approximated by
\begin{equation}
E(d^n) \cong E_0 + n \varepsilon_d + \frac{n(n-1)}{2}U_{vv} ,
\end{equation}
where $U_{vv}$ is the monopolar part of the valence-valence Coulomb interaction.  The charge transfer energy is then
\begin{equation}
\Delta \cong \varepsilon_d + n U_{vv} - \varepsilon_L   .
\end{equation}
However, $\Delta$ can differ from this value by several electronvolts
when considering the full multiplet hamiltonian.  In the Xclaim code,
the energy of the ligand-shell in Eq.~\ref{HybHam} is given by the
parameter $\varepsilon_L$ and not by $\Delta$.
In order to use the charge-transfer energy $\Delta$ as a parameter, we
need to calculate $\varepsilon_L$ for a given $\Delta$.
To do so, first we calculate the ground state energies
$E_{GS}(d^{n})$ and $E_{GS}(d^{n+1})$ for the $d^{n}$ and $d^{n+1}$
configurations of the metal center without taking into account the
ligand shell.  In the final calculation including ligands, the
energy level of the ligand shell is given in terms of $\Delta$ 
as $\varepsilon_L = E_{GS}(d^{n} )-E_{GS}(d^{n+1}) + \Delta$.

\section{Calculation of x-ray spectra}

For a one-photon process where the photon is absorbed, we can write the transition probability
using Fermi's golden rule 
\begin{equation}
\label{FGR}
I(\omega) = \sum_{f} | \langle f| T   |g\rangle |^2 \delta(E_f-E_g-\hbar\omega )     .
\end{equation}
$E_g$ and $E_f$ are the energies of the ground $|g\rangle$ and final states $|f\rangle$, respectively;
$\hbar\omega$ is the x-ray energy and $T$ is a
transition operator that connects the ground state to the
final states.  The particular form of $T$ depends on the x-ray
process that we are considering.
Eq.~(\ref{FGR}) can be reexpressed as a Green's function of the final state
\begin{equation}
\label{GreensFinal}
I(\omega)=  -\frac{1}{\pi}\mathrm{Im}   \langle g |T^{\dagger} 
\frac{1}{E_g+ \hbar\omega- H_f +i\Gamma}T|g \rangle                  ,
\end{equation}
where $\Gamma$ is the broadening due to the finite core-hole
lifetime.  In the calculation first the lowest-energy eigenstate $|g\rangle$ of the initial-state Hamiltonian is obtained and
the Green's function of the final-state Hamiltonian is calculated by using a continued
fraction expansion.

\subsection{X-ray Absorption Spectroscopy (XAS)}

In x-ray absorption a core electron is promoted to the valence
shell by an x-ray photon.  The transition operator in this case is
$T(E1)={\bm\epsilon}\cdot\mathbf{r}$ for dipolar transitions
and $T(E2)=({\bm \epsilon}\cdot\mathbf{r})({\bm k}\cdot\mathbf{r})$
for quadrupolar transitions.
${\bm\epsilon}$ is the x-ray polarization, $\mathbf{r}$
is the position operator and $\mathbf{k}$ is the propagation vector of the light.
The transition operators for dipolar and quadrupiolar transitions can be rewritten as spherical tensors
\begin{eqnarray}
\label{eqTsphe}
T(E1)&=& r \sum_q (-)^q \epsilon^{(1)}_{-q} C^{1}_q \nonumber \\
T(E2)&=& r^2 \sum_q (-)^q \sqrt{\frac{2}{3}}[\epsilon k]^{(2)}_{-q} C^{2}_q     ,
\end{eqnarray}
where the tensor product is defined as $[\epsilon k]^{(2)}_q = \sum \epsilon^{1}_{q'}  k^{1}_{q''} \langle 1 1 q' q'' | 2q\rangle$

The matrix elements of the spherical harmonics $C^{k}_q$ are given in~(\ref{MECkq}).
The radial matrix elements $r$ are constant for a given edge.
The transition operators for light with helicities  $\lambda=+1$ and $\lambda=-1$ correspond to 
setting $q=\pm 1$ in~(\ref{eqTsphe}).  Light with linear polarization along the $z$-axis corresponds to $q=0$.
Setting the radial matrix elements $r$ to unity and taking the light propagating along the $z$-axis $\mathbf{k}={\bf k}_z$
and linear polarization along the $x$ and $y$ axes the
transition operators for dipole and quadrupole transitions become
\begin{eqnarray}
T(E1,e_x) &=& \frac{1}{\sqrt{2}} (-C^{(1)}_{-1}+C^{(1)}_{1})   \nonumber \\
T(E1,e_y) &=& \frac{i}{\sqrt{2}} (C^{(1)}_{-1}+C^{(1)}_{1})   \nonumber \\
T(E2,k_z,e_x) &=& \frac{1}{   \sqrt{6}} (-C^{(2)}_{-1}+C^{(2)}_{1})   \nonumber \\
T(E2,k_z,e_y) &=& \frac{i}{   \sqrt{6}} ( C^{(2)}_{-1}+C^{(2)}_{1})    .
\end{eqnarray}

The program calculates linear and circular dichroism subtracting the
XAS for different polarizations .
X-ray magnetic circular dichroism  (XMCD) is defined as the difference between the spectra
for the incoming light with helicities  $\lambda=+1$ and $\lambda=-1$.

For XMCD, sum rules give a straightforward way to obtain
the orbital and spin magnetic moment from the integrated values of the measured
spectra~\cite{Thole1992,Carra1993}.
Similarly, it is possible to get the expectation value of the spin-orbit coupling
$\langle \sum_i \mathbf{l}_i \cdot \mathbf{s}_i\rangle$
from the branching ratio of the isotropic spectrum~\cite{Thole1988PRAp1943,Thole1988PRBp3158,vanderLaan1988PRL}.
The application of sum rules is not exempt from problems. The derivation of the sum rules with spin-dependent operators, such as the spin and the spin-orbit coupling, is based on the assumption that $j$ is a good quantum number at a particular spin-orbit split edge. 
However, mixing of the edges  occurs  as a result of other interactions, in particular the Coulomb core-valence interaction~\cite{Crocombette1996,vanderLaan2004PRL,Piamonteze2009}. 
For the spin sum rule, the presence of  the magnetic dipolar term $\langle T_z\rangle$ further complicates the determination of the value of the spin ~\cite{StohrKonig1995,Oguchi2004}. In addition to the calculation of the spectrum, Xclaim also calculates the expectation values of the most relevant tensor in the ground state. For a successful fit of the spectrum, these expectation values can provide a good estimate of these quantities in the material. Furthermore, they can serve as an additional check on the x-ray absorption sum rules.

\subsection{X-ray Photo-emission Spectroscopy (XPS)}

In X-ray Photo-emission Spectroscopy (XPS) the kinetic energy of an
emited electron is measured at a constant incident energy of the
x-rays.  We can calculate the core-level XPS with the cluster model similarly to the x-ray absorption
by using the annihilation of an electron in the core shell as the transition operator.  We can write the isotropic XPS
as
\begin{equation}
I^{\rm XPS}(\varepsilon) =  -\frac{1}{\pi}\mathrm{Im} \sum_{m\sigma}   \langle g|c^{\dagger}_{m\sigma} 
\frac{1}{E_g- \varepsilon - H_f+i\Gamma}c_{m\sigma}|g \rangle          ,
\end{equation}
where $c_{m\sigma}$ annihilates an electron with spin $\sigma$ in the $m$ orbital of the
core shell and $\varepsilon$ is the energy difference between the
photoelectron and the incident photon.

XPS is an ionizing proccess that produces large screening effects and
charge transfer satellites  appear accompanying the main peak of the spectra~\cite{vanderLaan1981}.
The spectral shape has a strong dependence on the magnitudes of the valence-valence
and core-valence monopolar part of the Coulomb interactions $U_{vv}$ and~$U_{cv}$.
When only considering the charge-transfer energy and the monopole parts of the Coulomb interaction, the energies of the
$|\underline{c}d^{n+1} \underline{L}^1\rangle$ and $|\underline{c}d^{n+2} \underline{L}^2 \rangle$ configurations relative to  $|\underline{c}d^{n}\rangle$
are $\Delta-U_{cv}$ and $2\Delta+U_{vv}-2U_{cv}$, respectively.
The spectroscopy final states that have a hole in the core shell and the Coulomb core-valence potential pull
down configurations with increasing number of electrons in the valence
shell.  For a configuration with $n$ electrons in the valence shell, its energy is decreased by $nU_{cv}$.
This effect usually produces a reordering of the final state configurations
and $|\underline{c}d^{n+1} \underline{L}^1\rangle$  appears below 
$|\underline{c}d^{n}\rangle$.  These two configurations are usually termed
the well-screened and the poorly screened final states.

\subsection{Photoemission and inverse photoemission}

We can also calculate the electron-removal
and electron-addition spectra, which can be observed in
valence photoemission spectroscopy (PES) and inverse photoemission
spectroscopy (IPES), respectively.~\cite{Zaanen1990}.
The angular integrated PES spectrum, given in terms of the difference between the 
energies of the photoelectron and incident photon $\varepsilon$ is
\begin{equation}
I^{PES} = -\frac{1}{\pi}\mathrm{Im} \sum_{m\sigma}   \langle g |d^{\dagger}_{m\sigma} 
\frac{1}{E_g- \varepsilon- H_f+i\Gamma}d_{m\sigma}|g \rangle   ,
\end{equation}
where $d_{m\sigma}$ annihilates an electron with spin $\sigma$ in the $m$ orbital of the
valence shell.

The IPES spectrum, as a function of the difference between the incident electron and
emitted photon $\varepsilon$ is calculated as
\begin{equation}
I^{IPES} = -\frac{1}{\pi}\mathrm{Im} \sum_{m\sigma}  \langle g |d_{m\sigma}
\frac{1}{E_g+\varepsilon- H_f+i\Gamma}d^{\dagger}_{m\sigma}|g \rangle   ,
\end{equation}
where $d^{\dagger}_{m\sigma}$ creates an electron with spin $\sigma$ in the $m$ orbital of the valence shell.

\section{Graphical interface}

\begin{figure*}
\begin{center}
\includegraphics[width=1.95\columnwidth]{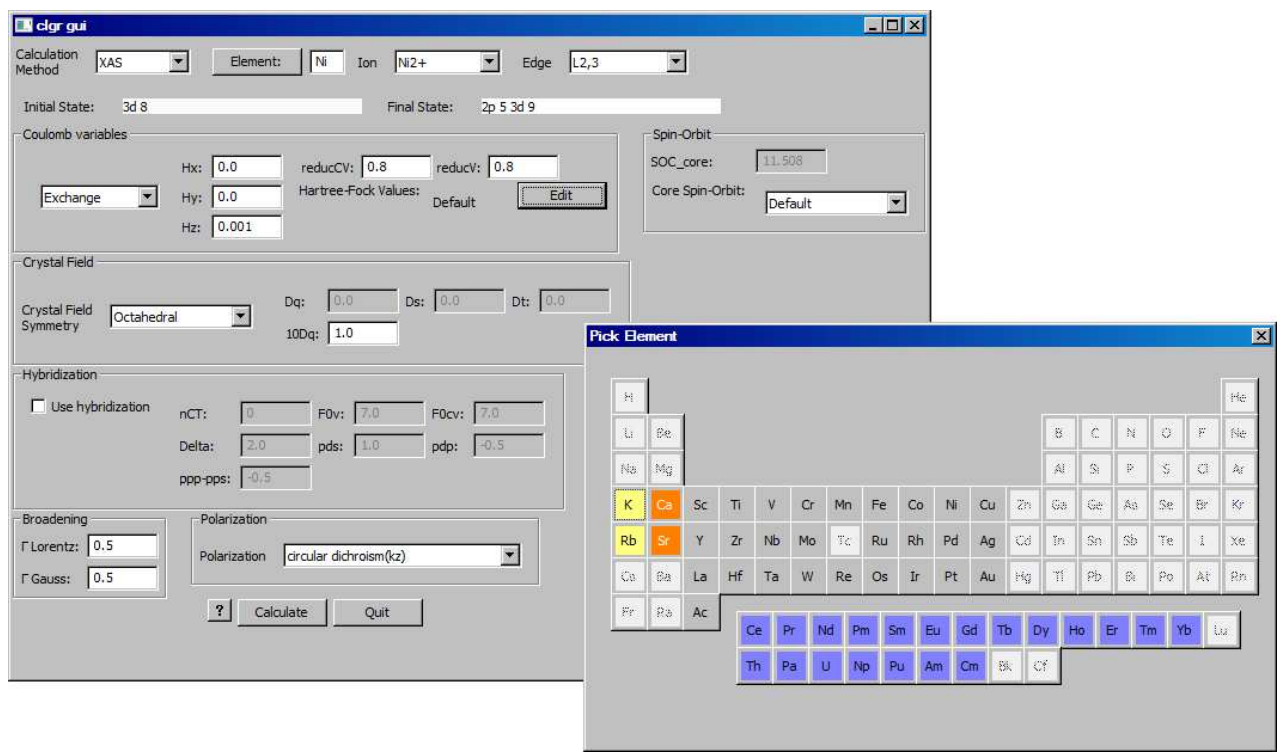}
\end{center}
\caption{\label{FigInput} Main input window and periodic table pop-up for the element selection.  The main window contains
the controls for setting up the main parameters for the spectra calculation divided into several areas: spectroscopy and ionization state, Coulomb and Spin orbit parameters, crystal field, hybridization, spectral broadenings and polarization.}  
\end{figure*}

When the program is started it displays a window
(Fig.~\ref{FigInput}) with entries for the
chemical element, ionization state and edge to be calculated,
as well as different Hamiltonian parameters.  Once the ion
and edge are chosen, the initial and final
state electronic configurations are automatically generated.
The program shows below the reduction values for the Slater
integrals for the Coulomb interactions within the valence shell, and
between core and valence.  The default reduction factor is 0.8.

\begin{figure}
\begin{center}
\includegraphics[width=0.48\columnwidth]{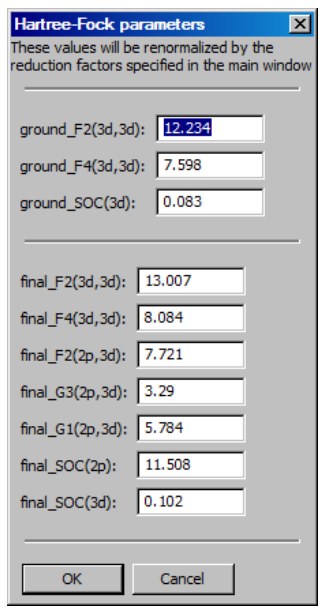}
\end{center}
\caption{\label{FigHF} Pop-up window showing the default Hartree-Fock parameters for the Slater
integrals $F^k$ and $G^k$ and the spin-orbit $\zeta$ of the core and valence shells. 
Customized values can be introduced in this window.}
\end{figure}

In addition to setting the reduction factors of the Coulomb interactions,
it is also possible to edit the full set of Slater
integrals and spin-orbit parameters for the initial and final states
by clicking in the button {\it Hartree-Fock values: Edit}.  
This opens a window (Fig.~\ref{FigHF}) where 
the Hartree-Fock parameters are separated into two different blocks
for the ground and final configurations.  The Slater integrals $F^k$, $G^k$
and spin-orbit parameters are labelled in terms of the different core and valence shells.
After clicking {\it Ok}, the values of the parameters are saved.
The values of the Slater integrals given in the window are
renormalized by the reduction factors specified in the main window.

The program allows the components of the magnetic field in the
$x,y,z$ directions to be specified.  
The two choices ({\it exchange} and {\it magnetic field})
mean that the field is acting on the spin moment
$\mathbf{S}$ or on the total magnetic moment of the ion
$\mathbf{\mu}=\mu_B(\mathbf{L}+2\mathbf{S})$.  The
exchange fields are given by setting $\mu_BH$ in units of eV
($\mu_B=5.79\cdot10^{-5}$~eV$\cdot$T$^{-1}$).

\begin{figure}
\begin{center}
\includegraphics[width=0.66\columnwidth]{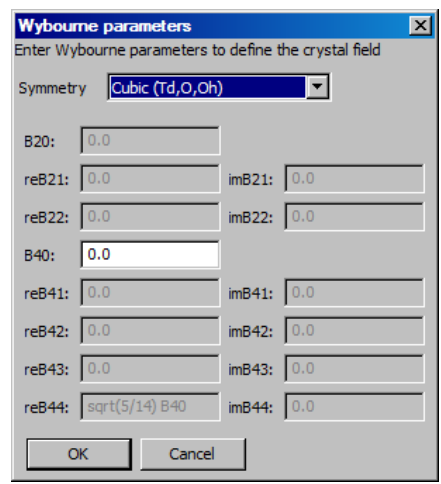}
\end{center}
\caption{\label{FigCF} Pop-up window for entering the crystal field when it is
defined in terms of Wybourne parameters $B_{kq}$.
The point group (cubic, tetragonal or hexagonal) can be selected from a pull-down list.
The point symmetry places constraints between the parameters and only the input boxes
for the independent parameters are active.
The input boxes for the parameters that are zero or constrained are inactive.
For the case of a general point group all the boxes are active and accept input.} 
\end{figure}

For setting the crystal field splitting,  one can select from a list of different
symmetries and parametrizations for the crystal field.
The main window allows to the values of the octahedral ($10Dq$) and
tetragonal ($Dq$, $Ds$ and $Dt$) crystal-field parameters.
For other crystal-field parametrizations, the parameters are set with pop-up dialog boxes.
One can specify the values of the energies of the different real $d$-orbitals, or in the case of a general point group,
it is possible to set the crystal field in terms of Wybourne parameters $B_{kq}$ (Fig.~\ref{FigCF}).
In the case of an $f$-valence shell the only way to set up the crystal field is
to specify Wybourne parameters.
The selection of {\it spherical} in the pull-down menu means that there is no crystal field term in the Hamiltonian.

The pop-up dialog box for setting the Wybourne parameters contains a
pull-down list to set the point-group symmetry.  For high symmetry
point groups (cubic, tetragonal, hexagonal) the program automatically
disables the input boxes for the parameters that arerequired to be zero by symmetry.
In the case of cubic octahedral symmetry, the only free parameters are
$B_{40}$ and $B_{60}$.  and the program automatically calculates $B_{44}$
and $B_{64}$.  Selecting point group symmetry {\it any}  in the pull-down list
means that there is no constraint in the Wybourne parameters and all
input boxes are activated.

The last group of parameters are related to the hybridization (implemented
for $O_h$ and $D_{4h}$ symmetry).  The first box is the maximum number of holes in
the ligand shell, this is the number of different electronic
configurations taken into account ($d^n$, $d^{n+1}\underline{L}$,
$d^{n+2}\underline{L}^2$...).  
The rest of the input boxes set the numerical values for the different
parameters involved in the hybridization Hamiltonian: charge-transfer
energy $\Delta$, isotropic coulomb interaction ($F^0$) for the valence
shell and for the attractive potential between the core-hole and 
valence electrons ($F^0_{cv}$), and the Slater-Koster parameters ($(pd\sigma)$,
$(pd\pi)$ and the difference $(pp\pi)-(pp\sigma)$)

From the parameters given, the program sets the Hamiltonian, and
calculates expectation values of quantum operators in the ground state
(energy, spin and orbital angular momentum and expected electronic
occupations of the valence and ligand shells) and  calculates the
spectra.  

%'isotropic','circular dichroism(kz)','linear dichroism(kx)','linear dichroism(ky)','linear dichroism(kz)'

%\subsection{Output windows}

\begin{figure}
\begin{center}
\includegraphics[width=0.99\columnwidth]{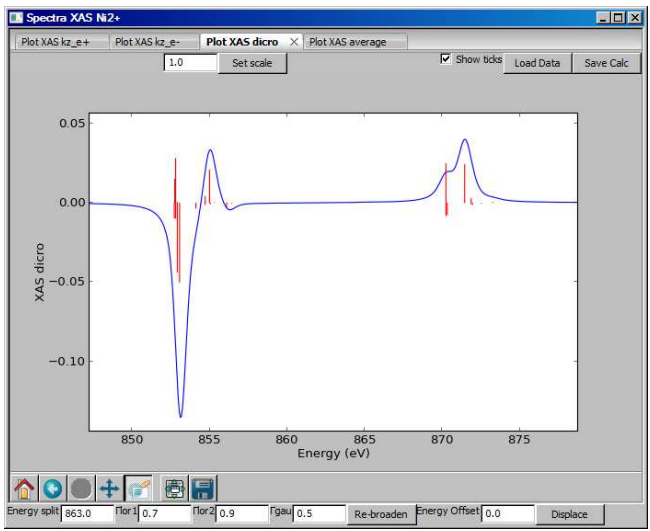}
\end{center}
\caption{\label{FigOutput} Window displaying the calculated x-ray absorption spectra.  Within the output window several tabs can be selected for viewing the
absorption curve for each polarization (left and right), their difference (dichroism) and the average of the two polarizations.
It possible to change the gaussian and lorentzian broadenings, set an arbitrary displacement in the x-axis for the calculated spectra and to load experimental
absorption curves for comparison with the calculated curve.} 
\end{figure}

The calculated spectra are shown in the output window (Fig.~\ref{FigOutput}).
For each of the polarizations calculated, results are placed in a
separate tab in the plot window.
In the case of dichroism, the difference (dichroism) and average
spectra for two polarizations are shown.
The multiplet model cannot account for the absolute positioning of the
absorption edge energy, so the program positions the edge according
to the values tabulated for the binding energies of the core-electrons
in different elements~\cite{TablasNISTurl}.
The calculated spectrum is displayed as poles (vertical bars) and also
convoluted with the input core-hole lifetime and experimental broadenings.
There are input boxes on the plot for setting the values of the Lorentzian and
Gaussian broadenings.  For core-hole spin-edges, it is possible to set
an energy-dependent Lorentzian broadening divided by an energy set by
the user. This is to account for possible differences in  core-hole lifetime broadening of the
two spin-orbit split edges due to the presence of additional Koster-Kronig processes at the edge at 
higher energy.
When the button {\it Rebroaden} is pressed, all polarization tabs in the
window are recalculated.
There are buttons on the plot window for loading experimental data to fit and to save the
calculation results to a file.

Another window shows the parameters used for the calculation 
and the expectation values of different physical
magnitudes in the ground state: number of holes in the ligand
and valence shells, 
the components of the total spin $\mathbf{S}$
and orbital angular momentum $\mathbf{L}$ given in units of $\hbar$, 
spin-orbit coupling $\sum_i\mathbf{l}_i \cdot \mathbf{s}_i$, 
and the magnetic dipole operator $T_z$ that appears as an additional term in the XMCD spin sum rule~\cite{Carra1993} (see appendix~\ref{SecDoubleTensors}).
For a $d$ valence shell the program also shows the individual occupation of the orbitals
$d_{3z^2-r^2}$, $d_{x^2-y^2}$, $d_{xy}$, $d_{yz}$, and $d_{zx}$.

\section{Conclusion}

We have presented a program for the calculation of core-hole
spectroscopy from a multiplet model, which gives
a good description of electron correlations and core-valence interaction.
The conventions used for the
model parameters are explained in terms of a general group theoretical
treatment.
The use of Wybourne parameters for the crystal field
allows the treatment of any point symmetry and makes
possible to fit x-ray spectra to a general crystal-field model.
Although our code does not include a first principles calculation of
crystal field parameters from the positioning of the ligands,
the ab-initio crystal fields constructed by codes such as Hilbert++~\cite{HilbertManual}
or MultiX~\cite{multixURL} can be mapped into the Wybourne parameter set.
Also, using this parametrizacion makes easy to relate the
crystal-field obtained from x-ray spectra with the results
derived by other techniques, such as inelastic neutron scattering.
The treatment of charge transfer in Xclaim allows for
valence-ligand charge transfer of an arbitrary number of electrons,
while the codes derived from Thole and Butler programs~\cite{Stavitski2010,missingURL} are limited to one-electron charge transfer.
The inclusion of several electrons charge transfer is important to
accurately simulate satellite peaks in x-ray photoemission
spectroscopy (XPS).
Applications include the fitting of x-ray spectra for
the determination of crystal fields parameters and ground state
configurations, 
sum rule error estimation or evaluating the effect in spectral
shapes of charge transfer effects.

\section{Acknowledgements}

We are thankful to D. Haskel, U. staub, and J. A. Blanco for useful discussions.
The periodic table was adapted from Robert Von Dreele's program pyFprime~\cite{pyFprime}.
This work was supported by the U. S. Department
of Energy (DOE), Office of Basic Energy Sciences, Division of Materials
Sciences and Engineering under Award No. DE-FG02-03ER46097, the
time-dependent x-ray spectroscopy collaboration as part of the
Computational Materials Science Network
(CMSCN) under grants DE-FG02-08ER46540  and DE-SC0007091, and NIU Institute for
Nanoscience, Engineering, and Technology. Work at Argonne National
Laboratory was supported by the U. S. DOE, Office of Science, Office of
Basic Energy Sciences, under contract No. DE-AC02-06CH11357.

\appendix

\section{Coupled tensor operators}
\label{SecDoubleTensors}

In this appendix we define the coupled tensor operators, which are
implemented in the program to calculate different quantum operators and
physical magnitudes.
For a shell with orbital $l$ and spin $s=\frac{1}{2}$ quantum numbers  
the unit tensor operator $w^{xyz}_\zeta$ is defined as~\cite{JuddBook,Thole1994PRB},
%%% EQS 14,17 in Thole1994PRB50
%
\begin{eqnarray}
{\rm w}^{xy}_{\xi\eta} &=& \sum_{mm'\sigma\sigma'}  (-1)^{l-m'} n^{-1}_{lx}n^{-1}_{sy}  \ThreeJSymbol{l}{x}{l}{-m'}{\xi}{m}    \nonumber \\
&&  (-1)^{s-\sigma'} \ThreeJSymbol{s}{y}{s}{-\sigma'}{\eta}{\sigma} c^{\dagger}_{m'\sigma'} c_{m\sigma}    
\end{eqnarray}
Where $a$ and $b$ are the unit tensor orbital and spin quantum numbers, with $-x\leq \xi \leq x$, $-y\leq \eta  \leq y$. The normalization factor $n_{lx}$ is defined as
\begin{equation}
  n_{lx}= \ThreeJSymbol{l}{x}{l}{-l}{0}{l}
\end{equation}
From the unit tensor operator we define the coupled tensor as,
\begin{eqnarray}
{\rm w}^{xyz}_q &=& \sum_{\xi\eta} (-1)^{x-\xi+y-\eta} n^{-1}_{xyz}  \ThreeJSymbol{x}{z}{y}{-\xi}{\zeta}{\eta}  {\rm w}^{xy}_{\xi,-\eta}  \nonumber \\
&&
\end{eqnarray}
with $-z \leq \zeta \leq z$.  $n_{xyz}$ is a normalization factor given by~\cite{EJPMvV}
\begin{equation}
n_{xyz}= \ThreeJSymbol{x}{y}{z}{0}{0}{0}
\end{equation}

%%%The matrix elements of the  normalized spherical harmonics  used to define the crystal-field potential in Eq.~(\ref{CFham}), can be obtained from the coupled double tensor operators by setting $y=0$ and $z=x$,
%%%\begin{eqnarray}
%%%{\rm C}^{k}_q &=& (-1)^{k} (2l+1) \ThreeJSymbol{l}{k}{l}{-l}{0}{l} \ThreeJSymbol{l}{k}{l}{0}{0}{0} {\rm w}^{k0k}_q \nonumber
%%%\end{eqnarray}

The double tensor operators $ {\rm w}^{xyz}_q$ are used to get the
ground state expectation values of physical observables: 
number of electrons in a shell $n_h =  {\rm w}^{000}_0$, 
total spin $\mathbf{S}=-s\mathbf{w}^{011}$ and orbital
angular momenta $\mathbf{L}=-l\mathbf{w}^{101}$, spin-orbit
coupling $\sum_i \mathbf{l}_i \cdot \mathbf{s}_i=lsw^{110}_0$, and
the magnetic dipole operator
$\mathbf{T}=\sum_i (\mathbf{s}_i-3\mathbf{r}_i(\mathbf{r}_i\cdot{\mathbf{s}_i})/r_i^2)=-\frac{l}{2l+3}\mathbf{w}^{211}_0$,
which is relevant for the analysis of XMCD, where
it appears as an additional term in the sum rule
used to determine the spin angular momentum~\cite{Carra1993}.

\newpage

%%%%%%%%%%%%%%%%%%%%%%%%%%%%%%%%%%%%%%%%%%%%%%%%%%%%%%%%%%%%%%%%%%%%
%%%%%%%%%%%%%%%%%%%%%%%%%%%%%%%%%%%%%%%%%%%%%%%%%%%%%%%%%%%%%%%%%%%%
%%%%%%%%%%%%%%%%%%%%%%%%%%%%%%%%%%%%%%%%%%%%%%%%%%%%%%%%%%%%%%%%%%%%
\bibliography{ManuscriptSPECGUI}
\end{document}